\pdfoutput=1
\documentclass[aps,prl,10pt,twocolumn,superscriptaddress,nofootinbib]{revtex4}

\usepackage{mathrsfs, amssymb, amsmath, amsfonts, txfonts, latexsym, graphicx}  
\usepackage[utf8]{inputenc}
	\usepackage[
      colorlinks=true,
      linkcolor=blue,
      urlcolor=cyan,
      filecolor=magenta,
      citecolor=red,
      pdfstartview=FitV,
      hyperfootnotes=false,
      pdftitle={Extended-BMS Anomalies and Flat Space Holography},
        pdfauthor={Laurent Baulieu, Luca Ciambelli, Tom Wetzstein},
        pdfsubject={Extended-BMS Anomalies and Flat Space Holography},
        pdfkeywords={Extended-BMS Anomalies and Flat Space Holography},
        bookmarksopen=true
      ]{hyperref}
\usepackage[all]{hypcap}
\usepackage{soul}

\usepackage{physics}

\makeatletter
\DeclareRobustCommand{\loplus}{\mathbin{\mathpalette\dog@lsemi{+}}}

\usepackage{float}
\usepackage{mathtools}

\newcommand{\dog@lsemi}[2]{\dog@semi{#1}{#2}{270,90}}
\newcommand{\dog@semi}[3]{%
  \begingroup
  \sbox\z@{$\m@th#1#2$}%
  \setlength{\unitlength}{\dimexpr\ht\z@+\dp\z@\relax}%
  \makebox[\wd\z@]{\raisebox{-\dp\z@}{%
    \begin{picture}(1,1)
    \linethickness{\variable@rule{#1}}
    \roundcap
    \put(0.5,0.5){\makebox(0,0){\raisebox{\dp\z@}{$\m@th#1#2$}}}
    \put(0.5,0.5){\arc[#3]{0.5}}
    \end{picture}%
  }}%
  \endgroup
}
\newcommand{\variable@rule}[1]{%
  \fontdimen8  
  \ifx#1\displaystyle\textfont3\else
    \ifx#1\textstyle\textfont3\else
      \ifx#1\scriptstyle\scriptfont3\else
        \scriptscriptfont3\relax
  \fi\fi\fi
}
\makeatother
\usepackage{tikz-cd}

\usetikzlibrary{arrows.meta,calc}

\usepackage{cleveref}
\usepackage{pict2e}
\usepackage{diagbox}

\newcommand\Lie{\pounds}

\newcommand{\beq}{\begin{eqnarray}}
\newcommand{\eeq}{\end{eqnarray}}
\newcommand{\beqn}{\begin{eqnarray}}
\newcommand{\eeqn}{\end{eqnarray}}
\newcommand{\pa}{\partial}

\newcommand{\cL}{{\cal L}}

\usepackage{pict2e}

\newcommand{\bz}{{\overline{z}}}

\newcommand{\rd}{\text{d}}

\newcommand{\chkM}{{\color{red} \,\checkmark\kern-5pt{}_{M}}}

\newcommand{\ee}{\end{equation}}
\newcommand{\bea}{\begin{eqnarray}}
\newcommand{\eea}{\end{eqnarray}}

\def\pa{\partial}

\usepackage{BOONDOX-cal}

\newenvironment{Align*}{\begin{equation*}
\begin{aligned}}
{\end{aligned}
\end{equation*}\par}

\usepackage{color}

\usepackage{layout}

\DeclareFontFamily{OT1}{rsfs}{}
\DeclareFontShape{OT1}{rsfs}{m}{n}{ <-7> rsfs5 <7-10> rsfs7 <10->rsfs10}{} 
\DeclareMathAlphabet{\mycal}{OT1}{rsfs}{m}{n}
\newcommand{\scri}{{\mycal I}}

\usepackage[utf8]{inputenc}
\usepackage{amsmath,amssymb,amsfonts,mathrsfs}
\usepackage{latexsym}
\usepackage{graphicx}

\usepackage[all]{hypcap}
\usepackage{cleveref}

\makeatletter
\DeclareRobustCommand{\loplus}{\mathbin{\mathpalette\dog@lsemi{+}}}

\usepackage{float}
\usepackage{mathtools}
\makeatother
\usepackage{tikz-cd}

\usetikzlibrary{arrows.meta,calc}

\usepackage{pict2e}
\usepackage{diagbox}

\usepackage{pict2e}

\def\pa{\partial}

\usepackage{BOONDOX-cal}

\def\demi{\frac{1}{2}}

\def\Lie{\mathcal{L}}

\def\={&=&}

\def\bea{\begin{eqnarray}}
\def\eea{\end{eqnarray}}

\def\w{\wedge}
\def\p{\pa_z}
\def\p0{\pa_0}
\def\bp{\pa_{\bar z}}

\def\bz{{\bar z}  }

\def\o {\omega}
\def\O {\Omega}

\def\E0 { \mathcal{E}^0  }

\def\Lie {\mathcal{L}}

\def \pbM  {   \begin{pmatrix}     }
\def \peM  {   \end{pmatrix}     }
\def \bM  {   \begin{matrix}     }
\def \eM  {   \end{matrix}     }

\usepackage{color}

\usepackage{layout}

\DeclareFontFamily{OT1}{rsfs}{}
\DeclareFontShape{OT1}{rsfs}{m}{n}{ <-7> rsfs5 <7-10> rsfs7 <10->rsfs10}{} 
\DeclareMathAlphabet{\mycal}{OT1}{rsfs}{m}{n}

\begin{document}

\setcounter{secnumdepth}{3}

\title{Extended-BMS Anomalies and Flat Space Holography}

\author{Laurent Baulieu}\email{baulieu@lpthe.jussieu.fr}
\affiliation{LPTHE, Sorbonne Universit\'e, CNRS, 
4 Place Jussieu, 75005 Paris, France}

\author{Luca Ciambelli}\email{ciambelli.luca@gmail.com}
\affiliation{Perimeter Institute for Theoretical Physics, 31 Caroline St. N., Waterloo ON, Canada, N2L 2Y5}

\author{Tom Wetzstein}\email{wetzstein.tom@gmail.com}
\affiliation{LPTHE, Sorbonne Universit\'e, CNRS, 
4 Place Jussieu, 75005 Paris, France}

\date{\today}

\begin{abstract}
We geometrically determine the BRST symmetry of a three-dimensional extended BMS field theory defined at the null boundary of asymptotically flat four-dimensional spaces. We then discuss its ghost number zero and one cocycles, independently of the four-dimensional bulk. The former determine a topological Lagrangian localized on the boundary of null infinity, while the latter classify the corresponding anomalies.
This classification reproduces the eBMS anomalies known from the bulk, providing a rare non-perturbative test of flat space holography. 
While supertranslations are anomaly-free, superrotations admit independent central charges. 
This theory can thus be dual to Einstein gravity only if these central charges are dictated by the bulk.
\end{abstract}

\maketitle


\section{Introduction}

A major turning point in the quest of quantizing gravity
is the discovery of its holographic nature \cite{Susskind:1994vu}, most concretely realized in the Anti-de Sitter (AdS)/Conformal Field Theory (CFT) correspondence \cite{Maldacena:1997re, Witten:1998qj}. 
A notable result in this context is the holographic derivation of the Weyl anomaly by Henningson--Skenderis \cite{Henningson:1998gx}, where boundary Weyl anomalies in the CFT were matched—and, in dimensions higher than four, even predicted—through a bulk asymptotic computation. 

The holographic description of gravity is revealed by the analysis of asymptotic symmetries, as precursory done by Brown--Henneaux \cite{Brown:1986nw}.
In $d+1$ AdS, the asymptotic symmetry group is $\text{SO}(2,d)$, which matches the global symmetry group of a CFT$_d$. Symmetry matching has been a guiding principle for exploring the holographic nature beyond AdS.

In this letter, we focus on spacetimes with vanishing cosmological constant, namely, asymptotically flat spacetimes. Here, the relevant boundary is null infinity $\scri^\pm$, and the asymptotic symmetries were famously analyzed by Bondi, van der Burg, Metzner, and Sachs, leading to the so-called BMS group \cite{Bondi:1962px,Sachs:1962wk,Sachs:1962zza}. 

This group is an infinite-dimensional extension of the Poincaré group, comprising a Lorentz sector along with supertranslations, which are translations that depend on the location on the celestial sphere. In $4$d, this group was further generalized in \cite{Barnich:2009se, Barnich:2010eb} by extending to locally well-defined vector fields on the celestial sphere, promoting the Lorentz group to include superrotations. This led to the definition of the extended BMS group (eBMS). Inspired by AdS/CFT, one is naturally led to seek an intrinsic eBMS field theory at $\scri^\pm$.

Flat space holography has largely been formulated in a bulk-to-boundary fashion.  
By contrast, an intrinsic definition of BMS field theories is still lacking, although important progresses have been made in \cite{Barnich:2010eb, Barnich:2014kra, Barnich:2015uva, Henneaux:2021yzg, deBoer:2023fnj, Salzer:2023jqv, Cotler:2024xhb, Nguyen:2025sqk}. 
This situation is unsurprising, since the null nature of asymptotic infinity poses significant conceptual and technical challenges.

We here make a step forward in the intrinsic understanding of the boundary field theory in flat space, by deriving its Lagrangian and anomaly structure, without any reference to a gravitational bulk. 
To classify them, we rely on the Becchi, Rouet, Stora, and Tyutin (BRST) symmetry and its geometric 
formulation. 
Although this approach may appear abstract, it proves to be an indispensable tool in our analysis.

We first determine the BRST symmetry of $\scri^\pm$.
Building on this, we show how the eBMS symmetry intrinsically arises through a suitable BRST-invariant gauge fixing.   
Next, we classify the corresponding eBMS-invariant Lagrangians that can be written as a wedge product, and anomalies, finding that the former are necessarily topological, while the latter appear only in the  superrotation sector.
In the spirit of \cite{Henningson:1998gx}, we then compare this boundary analysis to the bulk asymptotic structure, as studied in \cite{Baulieu:2023wqb, Baulieu:2024oql}. This gives a criteria for an eBMS field theory to be holographically dual to Einstein gravity. 
Our result provides non-perturbative support for flat-space holography, demonstrating a  bulk-boundary agreement and offering new insights into its codimension-$2$ structure.

We restrict our attention to the future null boundary $\scri^+$, as the analysis at $\scri^-$ is exactly the same.

\section{Geometry of Null Infinity} \label{sec_2}

Consider a $4$d asymptotically flat spacetime, in retarded Bondi coordinates $x^\mu=(r,x^a)=(r,u,x^A)=(r,u,z,\bz)$. In Bondi-Sachs gauge, the line element takes the form
\begin{align}
\begin{split}
\rd s^2 =&-2\rd u \rd r+g_{ab}(r,u,z,\bz)\rd x^a\rd x^b  \\
=&-\rd u^2-2\rd u \rd r +\frac{2m_B(u,z,\bz)}{r}\rd u^2\\
&+r^2 q_{AB}(z,\bz)\rd x^A\rd x^B+r C_{AB}(u,z,\bz) \rd x^A\rd x^B+\dots,
\end{split}
\end{align}
where $m_B$ denotes the Bondi mass and $C_{AB}$ the traceless asymptotic shear, which characterizes gravitational radiation.

Introducing $\rho=\frac1{r}$, the conformal boundary is at $\rho\to 0$, where the metric has an order-$2$ pole. The boundary metric of the conformally-compactified space is
\beq\label{ble}
\lim_{\rho\to 0}\Big(\rho^2\rd s^2\Big)=0\,\rd u^2+q_{AB}\rd x^A\rd x^B.
\eeq
This is the degenerate metric at future null infinity $\scri^+$, which underlies its \emph{Carrollian} structure.
This structure consists of the vector field $\ell=\ell^a\pa_a=\partial_u$ and the metric $q_{ab}=q_{AB}\delta^A_a\delta^B_b$, with $\ell^aq_{ab}=0$. The connection obeys \cite{Ashtekar:1981hw, Ashtekar:2024mme, Ashtekar:2024bpi, Ashtekar:2024stm}
\beq
D_aq_{bc}=0\qquad D_a\ell^b=0,
\eeq
and gives rise to an equivalence class spanned by the asymptotic shear \cite{Ashtekar:1981hw}. Given the analysis in \cite{Ciambelli:2023mir, Ciambelli:2025mex}, we can reformulate this statement introducing the Ehresmann connection $k_a$ \cite{Ciambelli:2019lap}, such that $\ell^a k_a=1$, and imposing $D_a k_b=\bar\theta_{ab}$, 
where $\bar\theta_{ab}$ is the radial expansion tensor:
\beq
\bar\theta_{ab}=\lim_{\rho\to 0}\frac12\cL_{\pa_\rho}\Big(\rho^2g_{ab}\Big)=\frac{C_{AB}}{2}\delta^A_a\delta^B_b.
\eeq
Therefore, the boundary connection coefficients are given by
\beq
\label{shear_connection}
\Gamma^a_{bc}=-\bar\theta_{bc}\ell^a\quad \Rightarrow\quad \Gamma^u_{AB}=-\frac{C_{AB}}{2}.
\eeq
This explicit expression for the Christoffel symbols in terms of the asymptotic shear is the same as in \cite{Nguyen:2022zgs}.

When considering a conformally-compactified manifold, the Weyl symmetry and the conformal class of Carrollian structures must be introduced, leading to the boundary conformal isometries
\beq
\label{conformal_isometries}
\Lie_\xi q_{ab} = 2 \lambda q_{ab}
\qquad
\Lie_\xi \ell^a = -\lambda \ell^a,
\eeq
with $\lambda = \demi \pa_A \xi^A$. This defines the conformal Carroll group, which is isomorphic to the BMS group  \cite{Duval:2014uva, Ciambelli:2019lap}. Allowing for locally well-defined vector fields on the $2$d sphere, the solution of \eqref{conformal_isometries} is
\beq
\label{EBMS_vector}
\xi=\xi^a\partial_a= \left(\alpha(z,\bz) + \frac{u}{2} \pa_A \xi^A\right)\partial_u+\xi^z(z)\partial_z+ \xi^\bz(\bz) \partial_{\bz} \, .
\eeq
This generates the eBMS symmetry group introduced in \cite{Barnich:2009se}. 

Note that the boundary structure discussed in this section is written in the adapted Carrollian frame \cite{Henneaux:1979vn,Bekaert:2015xua}. For a general Carrollian structure, this is reached in the associated frame bundle (see Appendix A of \cite{Ciambelli:2023xqk}) setting
\beq
\label{Carroll_tetrad}
    e^M_a = (k_a , e^I_a)\qquad  E^a_M = (\ell^a, E^a_I),
\eeq
where the frame bundle indices are $M=(u,I)=(u,z,\bz)$. 
We will use this formulation in the rest of the paper.

\section{Carroll BRST Symmetry}

We now present our new results, based on an intrinsic analysis of an eBMS field theory at $\scri^+$, independent of any dual bulk description. 
Its classical  symmetries are the diffeomorphisms $\text{Diff}(\scri^+)$ and the internal gauge symmetries: the $\text{U}(1)$ Weyl symmetry and the Lorentz--Carroll symmetry, 
the latter corresponding to the  Lorentz transformations preserving  the Carroll structure $(q_{ab},\ell^a)$.

To classify Lagrangians invariant under these symmetries, as well as their possible anomalies,
we employ the geometric BRST formalism introduced in \cite{Baulieu:1984pf,Baulieu:1985md}.
This opens the possibility for defining and computing physical cocycles, as will be done, which justifies the use of the geometric formalism.
We therefore work in the first-order formulation and extend the geometric structure of the previous section by adding the Lorentz--Carroll spin connection $\o^{MN}$.
This allows for a geometric description of the combined $\text{Diff}(\scri^+)$ 
$\times$ Weyl $\times$ Lorentz--Carroll  BRST symmetry through the construction of its graded differential operator $s$, which acts on both classical and ghost fields, combined in terms of the following geometrically meaningful bigraded one-forms:
\begin{align}
\label{bi-complex_one_forms}
    \tilde{e}^M = e^M + i_\xi e^M 
    \quad
    \hat{\o}^{MN} = \o^{MN} + \hat{\O}^{MN}  
    \quad 
    \hat{A} = A + \hat{\O} \, .
\end{align}
Here $i_\xi$ is the interior product along the $\text{Diff}(\scri^+)$ ghost field $\xi$, $\hat{\O}^{MN}$ are the ghosts for the Lorentz--Carroll symmetry, and $\hat{\O}$ is the $U(1)$ ghost of the Weyl gauge field $A$.

We then rely on the BRST bicomplex, with the  following unified spacetime and BRST symmetry exterior derivatives 
\begin{equation}
\label{d+s}
\hat\rd = \rd + \hat{s} \quad {\rm  and} \quad 
\tilde{\rd} = \rd + s = \hat{\rd} + \Lie_\xi \, .
\end{equation}
Here, $\rd = \rd x^a \pa_a$ is the spacetime exterior derivative on $\scri^+$ and $\hat{s} \equiv s - \Lie_\xi$ is the nilpotent BRST operator associated with the internal gauge symmetries. 
The conserved bigrading $(p,q)$ is crucial for expressing relevant equations, where $p$ represents the horizontal spacetime form degree and $q$ the vertical ghost number. Classical fields and the differential $\rd$ have ghost number zero, while the ghosts, $s$, $\Lie_\xi$ and $\hat{s}$ have ghost-number one, so \eqref{bi-complex_one_forms} 
and  \eqref{d+s} are consistent unifications.

Moreover, since one has $\hat{\rd} = \exp (-i_\xi) \, \tilde{\rd} \exp (i_\xi)$,  the nilpotency of the complete BRST operator $s$ is equivalent to that of $\hat{s}$. 
The consistency of the geometric BRST formalism guarantees that the closure and Jacobi relations of the  $\text{Diff}(\scri^+)$ $\times$ Weyl $\times$ Lorentz--Carroll gauge symmetry are equivalent to the nilpotency relations $(\rd+s)^2=(\rd+\hat s)^2=0$.

We are now ready to determine the internal gauge structure of an eBMS field theory and the action of the corresponding BRST operator on the fields \eqref{bi-complex_one_forms}. 
This is achieved by imposing horizontality conditions on the ghost-dependent torsion and curvatures, following the standard method of \cite{Baulieu:1984pf, Baulieu:1985md, Henneaux:1992ig, Barnich:2000zw}.

A distinctive feature of the Carrollian setting is that the frame bundle metric $\eta_{MN}$ is degenerate. We choose it to satisfy $\eta_{uN} = 0$, $\eta_{z \bz} = 1 = \eta_{\bz z}$ and $\eta_{z z}=0=\eta_{\bz \bz}$.
We then use $\hat\o^{M}{}_N=\hat\o^{MP}\eta_{PN}$ in the horizontality ($\hat{R}^{MN} = \hat{\rd} \hat{\o}^{MN} + \hat{\o}^M{}_P \w \hat{\o}^{PN} = R^{MN}$) and torsionless  ($\hat{T}^{M} = \hat{\rd} \hat{e}^{M} + \hat{\o}^{M}{}_N \w \hat{e}^N = T^{M} = 0$) conditions. By consistently adding the Weyl field $\hat{A}$, and using $\hat{e}^M \equiv \exp (-i_\xi) \tilde{e}^M = e^M$, we obtain for the torsion
\begin{align}
\begin{split}
\label{horizontality_diff_2}
    \hat{T}^z &= \hat{\rd} e^z   + ( \hat{A} + \hat{\o}^{z \bz} ) \w e^z = 0 \, , 
     \\
    \hat{T}^\bz &= \hat{\rd} e^\bz   + ( \hat{A} - \hat{\o}^{z \bz} ) \w e^\bz = 0 \, , 
     \\
    \hat{T}^u &= \hat{\rd} e^u + \hat{\o}^{u z} \w e^\bz + \hat{\o}^{u \bz} \w e^z + \hat{A} \w e^u = 0 \, , 
    \end{split}
\end{align}
and for the curvatures  
\begin{align} 
\begin{split}\label{horizontality_Lorentz_Weyl_2}
\hat{F} &= \hat{\rd} \hat{A} = F ,
     \\
    \hat{R}^{z\bz} &= \hat{\rd} \hat{\o}^{z \bz} = R^{z\bz} ,
     \\
      \hat{R}^{u z} &= \hat{\rd} \hat{\o}^{u z} + \hat{\o}^{z\bz} \w \hat{\o}^{u z}  = R^{u z} ,
       \\
      \hat{R}^{u \bz} &= \hat{\rd} \hat{\o}^{u \bz} - \hat{\o}^{z\bz} \w \hat{\o}^{u \bz}  = R^{u \bz}  .
      \end{split}
\end{align}

The ghost-number-one components of these equations determine the BRST transformation of the Carrollian geometric data, while the Bianchi identities of $\hat{T}^M, \ \hat{R}^{MN}$ and $\hat{F}$ ensure $\hat{s}^2=0$ on all fields so that the gauge algebra closes. 

From the ghost-number-two components of \eqref{horizontality_Lorentz_Weyl_2},
\beq
\label{ght}
\hat{s} \hat{\O} = 0 = \hat{s} \hat{\O}^{z \bz}\quad  \hat{s} \hat{\O}^{u z} = - \hat{\O}^{ z \bz } \hat{\O}^{u z} \quad \hat{s} \hat{\O}^{u \bz} =  \hat{\O}^{ z \bz } \hat{\O}^{u \bz},
\eeq
one deduces that the internal gauge group is $\text{U}(1)\times (\text{SO}(2) \ltimes \mathbb{R}^2 )$. Indeed, $\text{SO}(2) \ltimes \mathbb{R}^2$ preserves the Carroll structure and is the Inönü--Wigner contraction of the $3$-dimensional Lorentz group $\text{SO}(2,1)$. This is compatible with the fact that $\mathbb{R}^2$ generates the Carroll boost symmetry of \cite{Levy-Leblond:1965dsc}, with ghosts $\hat{\O}^{uz}$ and~$\hat{\O}^{u\bz}$.

In summary, the $\text{Diff}(\scri^+)$-covariant and nilpotent ghost transformations \eqref{ght} demonstrate that the relevant gauge group at $\scri^+$ is 
\begin{equation}
\label{Carroll_gauge_group}
\mathcal{G} = \text{U}(1)\times (\text{SO}(2) \ltimes \mathbb{R}^2 ) \times  \text{Diff}(\scri^+).
\end{equation}

\section{BRST Invariant Carroll Gauge Fixing}

We now identify the intrinsic eBMS symmetries as the residual symmetries of a suitably chosen BRST-invariant gauge fixing of \eqref{Carroll_gauge_group}.

Within $\text{Diff}(\scri^+)$, there is a special subgroup of diffeomorphisms, called Carrollian diffeomorphisms \cite{Ciambelli:2019lap}, which are automorphisms of the Carrollian fiber-bundle structure of $\scri^+ \simeq \mathbb{R} \times \Sigma^2$. These are  obtained by imposing  the conditions $\pa_u\xi^A=0$. These two conditions, together with  the seven degrees of freedom in \eqref{Carroll_gauge_group}, allow us to fix \eqref{Carroll_tetrad} as
\begin{equation}
\label{fully_GF_tetrad_2}
    e^M_{a} = \big( \delta^u_a , \delta^I_a \big) \qquad   E^a_{M} = \big( \delta_u^a, \delta^a_I \big) . 
\end{equation}
Inspired by the conformal gauge-fixing of the string worldsheet zweibein, we call this the  conformal Carroll gauge-fixing. This gauge choice is motivated by the adapted Carrollian frame of \cite{Henneaux:1979vn}. 

Requiring the consistency of the gauge \eqref{fully_GF_tetrad_2} under the BRST variation $s$ has two effects. First, it determines the internal ghosts in terms of the diffeomorphism ghosts. Second, it reduces the diffeomorphisms to a subgroup of the Carrollian diffeomorphisms.
Doing so leads to
\begin{eqnarray}
\begin{aligned}
    \label{ghosts_restriction_2}
   & \hat{\O} = \pa_u \xi^u  
      \qquad
    \hat{\O}^{u I} = \pa_{\bar{I}} \xi^u  
       \qquad
    \hat{\O}^{z \bz} =   \demi (\pa_z \xi^z - \pa_\bz \xi^\bz ) &
     \\
    &\xi^u = \alpha(z,\bz) + \frac{u}{2} \pa_A \xi^A  
      \qquad
    \xi^z = \xi (z)  
   \qquad
    \xi^\bz = \bar{\xi} (\bz) . &
    \end{aligned}
\end{eqnarray}
Remarkably, the second line of this equation is exactly the generator of the eBMS group, see \eqref{EBMS_vector}. While this group was derived from the bulk in the first section, we have here rediscovered it from an intrinsic boundary derivation, without any reference to an ambient space.

By using $s \xi^a = \xi^b \pa_b \xi^a$, we can construct the nilpotent BRST operator of the eBMS algebra \cite{Baulieu:2023wqb,Barnich:2017ubf}
\begin{align}
\label{s_BRST-BMS4}
s \alpha =\xi^A \pa_A \alpha + \frac{\alpha}{2} \pa_A \xi^A 
\quad
s\xi^z  = \xi^z  \pa_z \xi^z  
\quad
s \xi^\bz = \xi^\bz \pa_\bz \xi^\bz .
\end{align}

We know from \cite{Ashtekar:1981hw} that at $\scri^+$, the shear $C_{AB}$ is encoded in the connection. 
More directly, in the gauge \eqref{fully_GF_tetrad_2}, the torsionless conditions \eqref{horizontality_diff_2} indicate that $\o^{uz}_z + \o^{u\bz}_\bz$, $\o^{uz}_\bz$, and $\o^{u\bz}_z$ are left undetermined due to the degeneracy of the Carroll metric. As such they possibly are interacting fields.
These will be the only remaining physical degrees of freedom in the Carroll gauge. Indeed, using \eqref{shear_connection} and the relation between the spin connection and the Christoffel symbols, 
$\o^M_{a N } = e^M_d \Gamma^d_{ca} E^c_N + e^M_b \pa_a E^b_N$, we find that in the gauge \eqref{fully_GF_tetrad_2}  the only non-vanishing components of the spin connection are
\begin{align}
\label{gauge_fixed_SC_2}
    \o^{uI}   =   - \bar{\theta}^I{}_B dx^B = - \demi C^I{}_B \rd x^B .
\end{align} 
This relation relates the shear to the Carroll boost gauge fields. Since $C_{AB}$ is symmetric and traceless, we    have   $\o^{uz}_z = 0 = \o^{u\bz}_\bz$. 

We can now explicitly \textit{compute}, purely from the boundary perspective, the transformation of the shear under the BRST--eBMS operator \eqref{s_BRST-BMS4}. 
This is achieved by considering the ghost number one and $\rd  \bz$ component of the horizontality condition $\hat{R}^{uz} = R^{uz}$ \eqref{horizontality_Lorentz_Weyl_2},  in the gauge \eqref{fully_GF_tetrad_2}-\eqref{gauge_fixed_SC_2} and with  the residual ghosts \eqref{ghosts_restriction_2}.  We obtain
\begin{align}
\begin{split}
\label{full_shear_transformation}
    s C^z_{\ \bz} &= \xi^a \pa_a C^z_{\ \bz} + C^z_{\ \bz} \pa_\bz \xi^\bz -  2 \pa_\bz \hat{\O}^{u z} - \hat{\O}^{z \bz} C^z_{\ \bz} 
     \\
    &= \Big(  \xi^u \pa_u +  \xi^A \pa_A - \demi \pa_z \xi^z + \frac{3}{2} \bp \xi^{\bz}  \Big) C^z_{\ \bz} - 2 \pa_\bz^2 \xi^u  \, .
    \end{split}
\end{align}
The transformation of $C^\bz_{\ z}$ 
is as in  \eqref{full_shear_transformation}
with $z \leftrightarrow \bz$.
This novel intrinsic eBMS field-theoretical result correctly reproduces the known eBMS-transformation of the bulk gravitational shear \cite{Barnich:2010eb, Barnich:2011mi,Baulieu:2023wqb}. 

\section{EBMS Lagrangians and Anomalies}

We can now provide an intrinsic boundary classification scheme for eBMS-invariant Lagrangians and anomalies.
We work in the gauge \eqref{fully_GF_tetrad_2}, with the residual ghosts \eqref{ghosts_restriction_2} and the spin connection \eqref{gauge_fixed_SC_2},  which satisfies $\o^{z \bz}=0$. We moreover set $A$ to zero, as this can be done without affecting the symmetries. Under these conditions, we have $R^{z \bz} = 0 = F$. 

The BRST formalism then allows for a direct computation of invariant Lagrangians via the cohomology of $(3,0)$-forms and  of    anomalies as cohomological $(3,1)$-forms $\Delta^1_3$ \cite{Bonora:1982ve,Zumino:1983rz,Bonora:1983ff,Baulieu:1984iw}. 
In fact, as will be shown, the determination of $\Delta^1_3$ at $\scri^+$ will directly yield the  classification of invariant Lagrangians.
Therefore, we focus exclusively on $\Delta^1_3$. 

If the manifold under consideration has no boundary, the BRST anomaly consistency condition takes the form
\begin{equation}
\label{usual_WZ_condition}
    s \int \Delta^1_3 = 0 \quad \iff \quad s \Delta^1_3 = \rd \Delta^2_2 ,
\end{equation}
in which case the anomaly is defined modulo  $\rd-$ and $s-$exact terms. 
Null infinity, however, has non-trivial boundaries at $u=\pm \infty$. As a consequence, 
the equivalence  \eqref{usual_WZ_condition} between global and local conditions no longer holds, and one can rely only on the global one. In other words, eBMS anomalies at $\scri^+$ are classified by the $(3,1)$-forms $\Delta^1_3$ satisfying
\begin{equation}
\label{new_WZ_condition}
    s \int_{\scri^+} \Delta^1_3 = 0 \quad \text{with} \quad \Delta^1_3 \neq s \Delta^0_3\,, 
\end{equation}
where $s$ is the BRST--eBMS operator \eqref{s_BRST-BMS4}-\eqref{full_shear_transformation}. 
Indeed, the presence of boundaries implies that the $\rd u\, \pa_u$-exact part of a $\Delta^1_3=\rd \Delta^1_2$ satisfying \eqref{new_WZ_condition} can lead to non-trivial  anomalies. 

To solve \eqref{new_WZ_condition}, one must construct the most general bigraded 4-form $\hat{\Delta}_4$ built from the fields and curvatures, and impose \eqref{new_WZ_condition} to its $(3,1)$-component. 
Using \eqref{bi-complex_one_forms} and \eqref{horizontality_Lorentz_Weyl_2}, we find
a unique solution:\footnote{We report here only the final result. The full classification of all invariant polynomials  $\hat{\Delta}_4$ under the gauge group $\mathcal{G}$ defined in  \eqref{Carroll_gauge_group} is deferred to \cite{ToApp}.}
\begin{equation}
\label{Del_4}
    \hat{\Delta}_4 = \hat{R}^{uz} \hat{R}^{u \bz} - \hat{R}^{u \bz} \hat{R}^{u z} =   \hat{\rd} ( \hat{\o}^{uz} R^{u \bz} - \hat{\o}^{u \bz} R^{uz} ) = 0 . 
\end{equation}
The $(3,1)$-component of this equation reads  
\begin{equation}
\label{mixed_anomaly}
    \rd (  \hat{\O}^{uz} R^{u \bz} - \hat{\O}^{u \bz} R^{uz}  ) =   - ( s + \rd i_\xi  )(  \o^{uz} R^{u \bz} - \o^{u \bz} R^{uz} ) .
\end{equation}

\paragraph{\textbf{Lagrangian discussion}} Thanks to the specific cohomological problem at hand, we can directly derive the space of invariant Lagrangians.\footnote{We are classifying Lagrangians written as a wedge product. While, at least in YM-type gauge theories, all anomalies can be written in this way \cite{Dubois-Violette:1992idk}, there is no general theorem of this sort concerning Lagrangians.} 
Indeed, we deduce from \eqref{mixed_anomaly} that
\begin{equation}
\label{L_CS}
    L = \o^{uz} R^{u \bz} - \o^{u \bz} R^{uz} =
    \rd (\o^{u \bz} \o^{u z} ) = \frac{1}{4} \rd z \w \rd \bz \w  \rd u\,  \pa_u ( C^z_{\ \bz} C^\bz_{\ z}  ) 
\end{equation}
is the only eBMS-invariant  Lagrangian built out of the asymptotic shear.\footnote{If one studies  Lagrangians invariant under the larger group $\mathcal{G}$, the Chern--Simons type Lagrangian $L = \omega^{z \bar{z}} \rd \omega^{z \bar{z}}$ is also a valid candidate.}
It should be emphasized that this cohomological analysis does not aim to determine a complete microscopic boundary theory.  Additional boundary fields, describing for instance matter sectors or other ultraviolet completions of the bulk dynamics, may give rise to further invariant couplings involving those fields. Such model-dependent contributions lie beyond the scope of the present analysis.

The fact that this Lagrangian is $\rd$-exact, even if derived from an intrinsic analysis of a three-dimensional Carrollian field theory, echoes the proposed equivalence between the codimension-1 Carrollian holography and codimension-2 celestial holography programs \cite{Donnay:2022aba,Donnay:2022wvx}.

Note also that the Lagrangian \eqref{L_CS} precisely matches the corner counterterm structure arising from holographic renormalization in flat space, as derived in \cite{Compere:2020lrt,Odak:2022ndm}. This non-trivial correspondence will be analyzed in more detail in \cite{ToApp}.

Finally, it is worth observing that since the classical  Lagrangian \eqref{L_CS} is $\rd$-exact in a non trivial way, the method of \cite{Baulieu:1989rs} suggests, at least formally,  that it may define a local 
topological field theory  supported  on the boundaries of $\scri^+$. It will be shown in \cite{ToApp} that an eBMS-equivariant topological $Q$-exact BRST gauge-fixing imposes a specific $u$-dependence on the shear. As a result, one recovers the supertranslation effective Lagrangian of \cite{Nguyen:2021ydb}, whose two-point function captures all bulk gravitational infrared divergences.

\paragraph{\textbf{Anomaly discussion}} Since eq.~\eqref{new_WZ_condition} is defined up to $s$-exact terms, eq.~\eqref{mixed_anomaly} shows  that  the anomaly for the Carroll boosts $\hat{\O}^{uI}$ can always be rewritten as an anomaly for the diffeomorphisms $\xi$, compatible with the general arguments in \cite{Baulieu:1984pf,Bonora:1985cq,Baulieu:1989dg}. We thus confine our attention to the Carroll boost anomalies,
\begin{equation}
\label{mixed_anomaly2}
    \Delta^1_3 =   \rd (  \hat{\O}^{uz} R^{u \bz} - \hat{\O}^{u \bz} R^{uz}  ) \equiv \rd \Delta^1_2 .
\end{equation}

These potential anomalies are $\rd u \, \pa_u$-exact, establishing that all possible eBMS anomalies are supported at the boundaries of $\scri^+$. This is in line with the gravitational bulk interpretation of the anomaly \eqref{new_WZ_condition} as governing the infrared loop renormalization of the eBMS charges \cite{Baulieu:2024oql}, extending Noether's second theorem  to the anomaly structure.

Using the restricted eBMS ghosts \eqref{ghosts_restriction_2}, we decompose \eqref{mixed_anomaly2} into three independent potential anomalies:  $\Delta^1_{3,\alpha}$ for supertranslation and $\Delta^1_{3,\xi}, \, \Delta^1_{3,\bar{\xi}}$ for superrotations, proportional respectively to the ghost fields $\alpha$, $\xi$ and $\bar{\xi}$,
\begin{eqnarray}
\label{BMS_anomalies_decomposition}
\begin{aligned}
    \Delta^1_3 
    &=\demi  \rd z \w \rd \bz \w \rd u \, \pa_u \Big(   \,  (     \pa_z^2  C^z_{\ \bz} + \pa_\bz^2 C^\bz_{\ z} )\, \alpha 
     \\
    &+   \frac{u}{2} C^z_{\ \bz} \pa_z^3 \xi^z +   \frac{u}{2} C^\bz_{\ z} \pa_\bz^3 \xi^\bz \Big) \equiv   \Delta^1_{3,\alpha} + \Delta^1_{3,\xi} +  \Delta^1_{3,\bar{\xi}} \, .
    \end{aligned}
\end{eqnarray}
For these three contributions to be consistent and non-trivial anomalies, they must satisfy the $s$-cohomology, as given in~\eqref{new_WZ_condition}.

We start with supertranslations, thereby evaluating $s \int \Delta^1_{3,\alpha}$. Using \eqref{s_BRST-BMS4} and \eqref{full_shear_transformation}, we find the non-vanishing terms
\begin{equation}
   \int_{\partial \scri^+}  \rd z \w \rd \bz \, \Big( \frac{7}{2} \alpha \pa_z \xi^z \pa_z^2 + 2 \pa_z \alpha \pa_z \xi^z \pa_z  + \demi \pa_z^2 \alpha \pa_z \xi^z   \Big)C^z_{\ \bz}   \, + ...
\end{equation}
Therefore, the contribution $\Delta^1_{3,\alpha}$ does not satisfy \eqref{new_WZ_condition}: there are no anomalies in the supertranslation sector.

For superrotations, we must specify the asymptotic $u$-behavior of the shear.  Accounting for the loss of the peeling property \cite{Sahoo:2018lxl,Saha:2019tub,Geiller:2024ryw}, and working in a superrotation frame in which the vacuum news tensor vanishes, we impose the falloff\footnote{Choosing a shear falloff that depends on the vacuum news tensor and is consistent with superrotations does not alter our conclusions.} 
\begin{equation}
\label{def_C+-_2}
    C^z_{\ \bz}(u,z,\bz) \underset{u \to \pm \infty}{=} - 2 \pa_{\bz}^2 C_{\pm} (z,\bz) +  \frac{C^{L,\pm}(z,\bz)}{|u|} + o(|u|^{-1}).
\end{equation}
Then, using \eqref{s_BRST-BMS4}-\eqref{full_shear_transformation} and $\pa^2 \scri^+=0$, we compute 
\begin{equation}
\label{check_superrotation_anomaly}
    s \int_{\scri^+} \Delta^1_{3,\xi} =  \frac14 s \int_{\partial \scri^+} \rd z \w \rd \bz \ \pa^3_z \xi^z (C^{L,+} + C^{L,-}) = 0. 
\end{equation}
A similar conclusion pertains to $\Delta^1_{3,\bar{\xi}}$ for $z \leftrightarrow \bz$.

In a quantum field theory, the non-trivial superrotation cocycles $\Delta^1_{3,\xi} = \rd \Delta^1_{2,\xi}$ and  $\Delta^1_{3,\bar{\xi}} = \rd \Delta^1_{2,\bar{\xi}}$ signal a possible breaking of the corresponding Ward identities. Since the actual presence and magnitude of this breaking depends on the underlying dynamics, we introduce two model-dependent coefficients $c_{\xi}$ and $c_{\bar{\xi}}$ to quantify it.  
The breaking of the Ward identities is then governed by the combinations  $c_{\xi}\Delta^1_{3,\xi}$ and $c_{\bar{\xi}}\Delta^1_{3,\bar{\xi}}$.
Using the descent equations $s \Delta^1_{2,\bullet} = \rd \Delta^2_{1,\bullet}$, these same coefficients $c_\bullet$ multiply the ghost-number-2 cocycles $\Delta^2_{1,\bullet}$, which determine the central extensions in the eBMS algebra. In this precise sense, the coefficients $c_{\xi}$ and $c_{\bar{\xi}}$ are interpreted as central charges.

The superrotation anomalies of an eBMS field theory are thus characterized by the central charges $c_\xi$ and $c_{\bar{\xi}}$ and take the form
\begin{eqnarray}
\begin{aligned}
\label{boundary_superrotation_anomalies}
    &\Delta^1_{3,\xi}=\demi  \rd z \w \rd \bz \w \rd u \, \pa_u \Big( \frac{u}{2} C^z_{\ \bz} \pa_z^3 \xi^z \Big)\,,&  \\
    &\Delta^1_{3,\bar\xi}= \demi  \rd z \w \rd \bz \w \rd u \, \pa_u \Big(\frac{u}{2} C^\bz_{\ z} \pa_\bz^3 \xi^\bz\Big) \,. &
    \end{aligned}
\end{eqnarray} 

This is a central result of our analysis: the eBMS anomalies have been classified purely from a boundary perspective in the absence of matter. 
Supertranslations are found to be anomaly-free, whereas superrotations can be anomalous, each governed by an independent central charge. We now turn to comparing these results with the corresponding gravitational analysis in the bulk.

\section{Holographic Matching}

Let us recall that the first holographic test is based on symmetries: the bulk asymptotic symmetries are dual to the global symmetries of the boundary field theory. With suitable boundary conditions, flat space asymptotic symmetries are the eBMS symmetries \cite{Bondi:1962px, Sachs:1962wk, Sachs:1962zza, Barnich:2009se, Barnich:2010eb, Strominger:2013jfa}. Thus, the boundary $\scri^+$ hosts  an eBMS field theory.

Under which conditions is this eBMS field theory dual to gravity in the bulk of an asymptotically-flat spacetime? First, the absence of supertranslation anomaly is dual to the bulk statement that Weinberg's soft graviton theorem \cite{Weinberg:1965nx} is exact at tree level, and thus does not receive quantum corrections.

Then, analogously to how the values of the CFT central charges are determined by the AdS bulk \cite{Henningson:1998gx}, we predict the values of the superrotation central charges $c_\xi$ and $c_{\bar{\xi}}$ through a holographic matching. 

The anomaly $\Delta^1_{3,\xi}$ was previously found in the bulk in \cite{Baulieu:2024oql,Choi:2024ygx,Choi:2024ajz}, where it is interpreted as the 1-loop infrared-renormalized superrotation soft charge $Q_{\xi,\text{soft}}^{\log} = c_{\xi,\text{bulk}}  \int_{\scri^+} \Delta^1_{3,\xi} $. This charge, once inserted in the Ward identity $\bra{\text{out}}[Q_{\xi,\text{soft}}^{\log} , S] \ket{\text{in}} = 0$ for the bulk $S$-matrix, inserts a soft graviton and isolates the $\ln{\o}$ term in the soft expansion in $\o$ of the logarithmic soft graviton theorem \cite{Choi:2024ajz}.\footnote{Our analysis cannot capture the effect of adding massive particles in the bulk, and thus the anomaly responsible for the 1-loop correction to the kinematic factor of the logarithmic soft graviton theorem \cite{Sahoo:2018lxl}. Including this
would require analyzing the eBMS anomalies at $\scri^+ \cup i^+$.} Its explicit expression is given by 
\begin{equation}
    \label{bulk_anomaly}
    Q_{\xi,\text{soft}}^{\log} =  - \frac{2}{\kappa^2} \int_{\scri^+} \rd u \w \rd z \w \rd \bz  \,  \pa_u (u^2 \pa_u C^z_{\ \bz}) \pa^3_z \xi^z    ,
\end{equation}
where $\kappa^2= 32 \pi G$ and $G$ is the gravitational constant. The charge $Q_{\bar{\xi},\text{soft}}^{\log}$ exhibits the same numerical prefactor. 

Using the $u$-falloff \eqref{def_C+-_2}, the bulk result \eqref{bulk_anomaly} precisely matches the boundary superrotation anomalies $c_{\xi}\Delta^1_{3,\xi}$ and $c_{\bar{\xi}}\Delta^1_{3,\bar{\xi}}$ given in  \eqref{boundary_superrotation_anomalies}, provided the eBMS dual field theory has
\begin{equation}
\label{holographic_matching}
    \boxed{ c_{\xi} = \frac{1}{4 \pi G}\qquad c_{\bar\xi} = \frac{1}{4 \pi G} } \ .
\end{equation}

This result deserves some comments. First, it indicates that the right and left movers of the celestial CFT have the same central charge, dictated by the bulk: this is the flat-space analog of the $a=c$ theorem in AdS/CFT \cite{Henningson:1998gx}. Secondly, since both the central charges $(c_{\xi},c_{\bar\xi})$ and  the cocycles $(\Delta^1_{3,\xi},\Delta^1_{3,\bar{\xi}})$ in \eqref{boundary_superrotation_anomalies} are dimensionful, the resulting central extensions of the eBMS algebra—controlled by $c_\bullet \Delta^2_{1,\bullet}$—are field-dependent and dimensionless. This is consistent with the interpretation 
of the eBMS anomalies as renormalized soft charges.  Lastly, the connection between these central extensions and previously studied ones \cite{Barnich:2011mi, Fotopoulos:2019vac, Banerjee:2022wht, Rignon-Bret:2024gcx} is worth exploring. A preliminary study indicates that our central extensions may alter the factorization of the two superrotation sectors, due to the presence of radiation. Indeed, once integrated, the cocycles \eqref{boundary_superrotation_anomalies} vanish in the absence of radiation.

\section{Final Words}

In this letter, we explored $3$d eBMS field theories through a BRST framework, independently of a gravitational bulk. By consistently incorporating the shear via the boundary connection and the geometry of the frame bundle,  we identified the gauge group ${\cal G}$ and derived the associated BRST $s$-transformations from first principles. We introduced the conformal Carroll gauge, which is naturally preserved by the eBMS group, and obtained the $s$-transformation of the boundary connection \eqref{full_shear_transformation}, which reproduces the bulk transformation of the asymptotic shear solely from the boundary.

We then examined invariant Lagrangians and anomalies. Both turned out to be $\rd$-exact, consistent with the theory effectively localizing to the boundary of $\scri^+$. We then found that while supertranslations remain anomaly-free, superrotations carry independent anomalies, each with its own central charge. By comparing with the bulk, we showed that a consistent holographic picture emerges only when these central charges match, as given in eq.~\eqref{holographic_matching}. This is a flat-space analogue of the famous AdS/CFT result \cite{Henningson:1998gx}.

These results support the idea of a boundary theory with two-dimensional characteristics, and constitute one of the few non-perturbative  confirmation of flat-space holography. Therefore, our findings align naturally with celestial holography: while a CFT$_3$ dual to AdS$_4$ has no central charges, our results suggest an unusual CFT$_2$ dual could better capture the physics of four-dimensional flat space, highlighting subtle challenges in taking the flat limit of AdS$_4$.

Our results suggest several directions for future investigation. In particular, it would be interesting to further clarify the role of superrotation anomalies in the boundary theory and their realization in Carrollian/eBMS field theories. Exploring the consequences of their cancellation through suitable Liouville sectors, as well as the structure of the associated boundary current algebra, also appears worthwhile. 

\vspace{0.25cm}

\paragraph{Acknowledgements}

LB and TW thank Marc Bellon, Sangmin Choi and Andrea Puhm for discussions. 
LC is grateful to Glenn Barnich, Rob Leigh, Sruthi Naranayan, and Simone Speziale for discussions. Research at Perimeter Institute is supported in part by the Government of Canada through the Department of Innovation, Science and Economic Development Canada and by the Province of Ontario through the Ministry of Colleges and Universities. LC is supported by the Simons Collaboration on Celestial Holography. TW is grateful to LC for hospitality at Perimeter during the initial stages of this work.

\bibliographystyle{uiuchept}
\bibliography{biblio_letter.bib}


\end{document}